\documentclass[
aip,
jcp,%
 amsmath,amssymb,
 reprint,%
]{revtex4-1}

\usepackage{graphicx}
\usepackage{dcolumn}
\usepackage{bm}
\usepackage{CJK}
\usepackage{booktabs}
\usepackage{multirow}					
\usepackage{array}
\usepackage{footnote}

\begin{document}

\preprint{AIP/123-QED} 

\title{Machine learning enhanced global optimization by clustering local environments to enable bundled atomic energies}
\author{S\o{}ren A. Meldgaard}

\affiliation{ 
Department of Physics and Astronomy and Interdisciplinary Nanoscience Center (iNANO), Aarhus University, 8000, Aarhus, Denmark.
}
\author{Esben L. Kolsbjerg}
\affiliation{ 
Department of Physics and Astronomy and Interdisciplinary Nanoscience Center (iNANO), Aarhus University, 8000, Aarhus, Denmark.
}
\author{Bj\o{}rk Hammer}
\email{hammer@phys.au.dk.}
\affiliation{ 
Department of Physics and Astronomy and Interdisciplinary Nanoscience Center (iNANO), Aarhus University, 8000, Aarhus, Denmark.
}
 
\date{\today}

\begin{abstract}
We show how to speed up global optimization of molecular structures using machine learning methods. To represent the molecular structures
we introduce the \textit{auto-bag} feature vector that combines: i) a
local feature vector for each atom, ii) an unsupervised clustering of
such feature vectors for many atoms across several structures, and
iii) a count for a given structure of how many times each cluster is
represented. During subsequent global optimization searches, accumulated
structure-energy relations of relaxed structural candidates are used
to assign local energies to each atom using supervised learning.
Specifically, the local energies follow from assigning energies to
each cluster of local feature vectors and demanding the sum of local
energies to amount to the structural energies in the least squares
sense. The usefulness of the method is demonstrated in basin hopping
searches for 19-atom structures described by single- or double-well
Lennard-Jones type potentials and for 24 atom carbon structures
described by density functional theory (DFT). In all cases, utilizing
the local energy information derived on-the-fly enhances the rate at
which the global minimum energy structure is found.
\end{abstract}

\maketitle

\section{\label{sec:introduction}Introduction}
The field of atomic-scale structure search is crucial in a wide span of disciplines
ranging from catalysis over material science to molecular biology.
For an efficient search for the structural global
minimum in an energy landscape of many dimensions it requires
optimization techniques of global character. Such methods include
random search \cite{RS}, basin hopping (BH) \cite{BH},
and evolutionary algorithms \cite{GA1, GA2, GA3, GA4, GA5, GA6,
  GA7, GA9, GA8}. Common for these methods are that they facilitate an increased
exploration of configuration space compared to e.g.\ molecular dynamics
driven searches, that excel on the exploitational search in already identified
funnels of the energy landscape. 

First principles methods, that take no input from experiment and
involve no empirical parameters, are the methods of choice for
atomic-scale structure search.  Notably, density functional theory
(DFT) is being used for structure optimization, as it has proven
highly accurate in reproducing observed structures,
rationalizing experimental observations where the underlying
structures were elusive \cite{tio2-ridge,sno2}, and in some cases even predicting new
structures \cite{b-cluster}. The high
accuracy of DFT builds on solving the quantum mechanical problem of
single-particle Hamiltonians for electrons in effective potentials and
on representing the electrons through a self-consistently found
all-electron spatial density. These elements of DFT must be attended for
each energy and force evaluation during structural searches and make
the computational method highly demanding.

Recently, machine learning (ML) procedures have been introduced to map
the atomic interactions from the elaborate DFT framework to predefined functional expressions such as force fields \cite{FF}, or to general fitting frameworks such as linear regression \cite{MM}, neural networks \cite{BP, DPMD, IBE, NN, DTNN, SchNet, AAP} and kernel based regression \cite{CM1, KRR2, GAP_orig, GAP1, GO, NEBGP}.

These ML-based methods demonstrate tremendous speed-ups for predicting DFT energies and forces at only
moderate loss of accuracy. This has been utilized in global optimization by consulting the ML-model instead of expensive DFT or higher order calculations allowing for searching at a fraction of the original cost. Furthermore, protocols have been established in which the accuracy of the ML-model is monitored while the ML-based potentials are being used in order to capture failure and re-adjust the ML models \cite{AEA1, AEA2, NNEA, GAP_sampling}. The use of ML has thus greatly reduced the computational cost of exploring configurational space allowing for more efficient structure searches.

Another benefit is that with the introduction of ML methods, a local energy concept often
emerges \cite{BP,GAP_orig, NN, GRAPE}, which is not present in DFT unless extra measures are taken
\cite{Dallas_Trinkle}. A local energy concept is useful in the context
of structure optimization as it opens for directing the structural
search more efficiently by focusing on improving the high energy
regions of structures while preserving structural arrangements in low
energy regions. This has been shown in previous work employing predefined descriptors and an evolutionary algorithm \cite{EAR, xin}.

In the present work, we add to the current developments in the use of
ML methods in chemical physics by introducing a simple means of
representing atomic structures with the \textit{auto-bag}
feature vector and by formulating a simple regression framework, that
enables the extraction of average local atomic energies based on grouping local environments. The method is
demonstrated to work with very little input data at the DFT level,
which potentially makes it interesting when probing new systems, where
DFT data is scarce. Local energies are then used to direct the search towards perturbing unstable regions of the structure.

\begin{figure*}[tb]
  \centering
   \includegraphics[width=2.0\columnwidth]{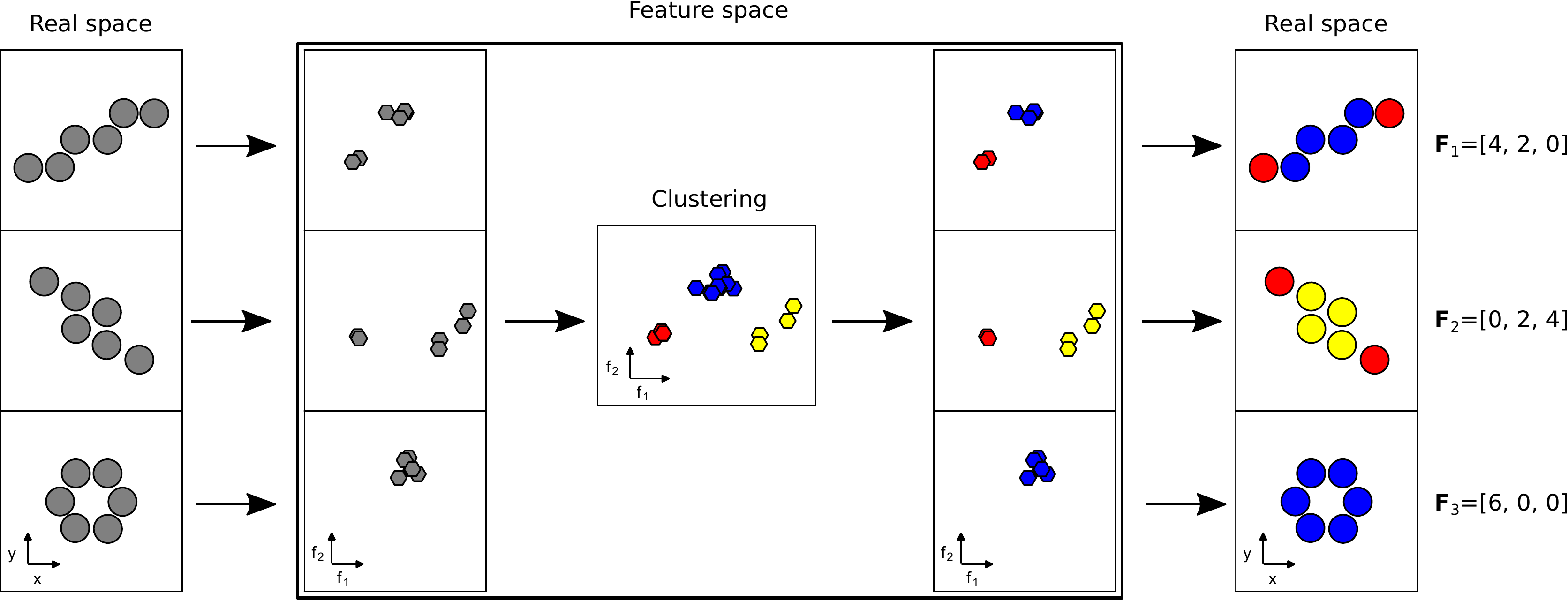}
   \caption{Schematic illustration of the method for extracting the
     \textit{auto-bag} feature vector. From left to right: structures
     with atoms (grey disks) in various local environments.  Atomic
     feature vectors (grey hexagons) are extracted. The feature
     vectors for all structures are clustered (colored hexagons). Each
     structure may subsequently be illustrated either in feature space
     (structure-wise plots of colored hexagons) or in real space
     (colored disks). By counting the abundance of members of each
     cluster in a given structure, $I$, the \textit{auto-bag} feature
     vector, $\mathbf{F}_I$, is obtained.}
  \label{fig:schematic}
\end{figure*}

The paper starts by describing the method in the
context of a BH-enabled structural search for the optimum
2D structure using only classical Lennard-Jones (LJ) potentials for the atomic
interactions. The BH method and the LJ potentials are sufficiently simple,
that students and researchers with no prior training in the field 
may readily code these elements and reproduce the
presented results. Also, the use of the LJ potentials has the added benefit
that the ML-enabled local energies can be directly compared to the
true LJ-based local energies, which would not be possible in a DFT-framework.
The paper proceeds with a demonstration of the applicability of method when
used in conjunction with DFT in a search for the optimum 2D-cluster shape
of a 24 carbon atom structure.

\section{\label{sec:method}Method}

\subsection{Representation}
When ML methods are introduced, the first concern is a
proper representation of the data on which the learning is
made. Representation is highly domain specific. In text processing,
the \textit{bag-of-words} vector, that counts the occurrence of known
words and neglects grammar and word order, is often used to map an
entire text to a simple vector of integers. Likewise, performing
customer segmentation in a retail or web shop, a customer (as
identified by a credit card or a login) may be represented by his or
her historical spending in various predefined product categories in
what we may dub a \textit{bag-of-spendings} vector.

In the chemical physics domain, structures are naturally described by
a list of atomic identities and corresponding cartesian coordinates,
and -- for space filling matter -- by the super cell vectors. However,
such a representation is not adequate for ML \cite{REP}, since it is
not invariant to translations, rotations, or permutation of identical
atoms -- operations that do otherwise leave the physical properties of
the compounds unchanged.

Several representations have been proposed that deal with this
deficit.  One of the simplest such representations is one in which all interatomic separations
are evaluated, sorted in ascending order, and kept in "bags" of AA,
AB, BB, $\dots$ bond-type, where A, B, $\dots$ are the atomic identities. The resulting
\textit{bag-of-bonds} vector thus represents the entire
structure \cite{BOB}. Several other global descriptors have been proposed,
including the fingerprint feature vector \cite{FP} and the coulomb
matrix representation \cite{CM1, CM2}. Presently, much work in the field relies on local feature
vectors such as the Behler-Parrinello symmetry functions \cite{BP}, smooth overlap of atomic positions (SOAP) \cite{SOAP} and others \cite{GRAPE, AF1, AF2}. These
feature vectors describe the local environment of
each atom in a given structure and represent entire structures by the collection of such
atomic feature vectors.

To keep the complexity of the method introduced in this work at a
minimum and to construct a method that requires only little data to
provide useful predictions, we propose the \textit{auto-bag} feature vector
representation, which will be detailed in the following. The method
has two elements, (i) the automated identification of bags of atomic
environments, and (ii) the representation of a structure as the count
of the occurrence of those environments in that given structure.

Fig.\ \ref{fig:schematic} presents the build-up of the
\textit{auto-bag} feature vector. Its starts by evaluating atomic
feature vectors for each atom in a collection of structures. Next, the
feature vectors are grouped in "bags", and finally for each structure
the abundance of group members may be counted and collected as a
vector of integers. The groups act as the classification in line with
the "known words", the "predefined product categories", and the "AA, AB,
and BB bond-types" in the bag-of-words, bag-of-spendings, and bag-of-bonds
methods discussed above. However, by using a ML technique, \textit{clustering}, to identify the groups, they need not be
predefined, and hence the naming: the \textit{auto-bag} feature vector.

In Fig.\ \ref{fig:schematic}, the \textit{local} feature vector describing each atomic
environment is considered two-dimensional for illustration purposes, but it may in fact
be chosen as simple as a single number, e.g.:
\begin{equation}
  \label{eq:1dsym}
  f_i(\mathbf{r}_1, \dots , \mathbf{r}_N) = f_i^\alpha
\end{equation}
where $N$ is the number of atoms and $f_i^\alpha$ is a function describing
the local density around atom $i$ within some cutoff distance ($\alpha$ being a label). A
simple example of such a local feature is the radial symmetry function as given by Eq.~(\ref{eq:symfuncR})
in the appendix.  However, a radial symmetry function will not
uniquely define the local environment. It will for instance not be able to differentiate between
configurations with two atoms located a distance $r$ from atom $i$ but
with different bond angles. To encapsulate angular information the
local feature vector can be expanded to
\begin{equation}
  \label{eq:2dsym}
  \mathbf{f}_i(\mathbf{r}_1, \dots , \mathbf{r}_N) =
  \begin{bmatrix}
    f_i^\alpha \\
    f_i^\beta
  \end{bmatrix}
\end{equation}
where $f_i^\beta$ is an angular symmetry function given by Eq.~(\ref{eq:symfuncA}) in the appendix ($\beta$ being a label). For an even more detailed description of the local environment several radial and angular symmetry functions can be used to yield a higher dimensional local feature vector as described in the appendix. 

The grouping of local feature vectors illustrated in the middle of
Fig.\ \ref{fig:schematic} must be done using an unbiased method in
order for the method to work autonomously.  This is possible using
unsupervised ML techniques known as clustering
methods. A clustering method acts to identify relations and propose a
categorization of data without prior definition of the categories --
hence the adjective "unsupervised". Many clustering schemes have been
proposed and some have even been demonstrated to enable speed-up of
structural search \cite{EAC, knud}. In the present context, we have found
the simple clustering method, the K-means algorithm \cite{kmeans}, sufficient to
fulfill the needs of establishing the auto-bag feature
vector. K-means takes the number of desired clusters as input and is
non-deterministic in that its random initialization may cause new
results in repeated uses on the same data.
This property is not a problem in
connection with structure optimization where a certain level of
stochastic behavior may even be desirable \cite{knud}. 
To stabilize the method
using K-means, we did, however, use the K-means++ initialization \cite{kmeanspp} to achieve a reasonable local optimum and prevent empty clusters.

Once the clustering is done, a structure will be described by a global feature vector:
\begin{equation}
  \label{eq:globfeature}
  \mathbf{F}(\mathbf{f}_1, \dots, \mathbf{f}_N) = [n_1, \dots , n_c, \dots, n_C] ,
\end{equation}
where $n_c$ is the number of atoms in cluster $c$ with $C$ clusters in total. The global
feature vector respects invariance to permutation of identical atoms, and inherits translational and rotational symmetry from
the local feature vectors.

\subsection{Local energies}
We now propose to parameterize the system energy as
\begin{equation}
  \label{eq:locenergy}
  E(\mathbf{r}_1, \dots, \mathbf{r}_N) = \sum_{i = 1}^{N} \epsilon(\mathbf{f}_i) \approx \sum_{i = 1}^{N} \varepsilon_{c(i)} = \sum_{c = 1}^C n_c \varepsilon_c, 
\end{equation}
where $\epsilon(\mathbf{f}_i)$ is the local energy of atom $i$ as defined by its local feature vector, $\varepsilon_c$ is a common local energy assigned to all
members of cluster $c$, and $c(i)$ is the cluster index for atom
$i$. We note that writing the total energy of a structure as a sum of local energies is an approximation. By bundling the local energies, i.e.~forcing them to
be identical for all members of a given cluster, the approximation becomes
even more severe. However, choosing such bundled cluster-wise local energies means that the method has fewer free
parameters and that extraction of meaningful energies can be done with
the simple method that follows below.

Introducing an index, $I$, that enumerates the structures we have the
relation:
\begin{equation}
  \label{eq:totalEnergy}
  E_I = \sum_{c = 1}^C n_{Ic} \varepsilon_c
\end{equation}
defining the unknown local energies $\varepsilon_c$ as a function of the global feature-energy relations, $(\mathbf{F}_I,E_I)=([n_{I1},\dots, n_{Ic}, \dots, n_{IC}],E_I)$.
By observing multiple structures, a matrix problem emerges:
\begin{equation}
  \label{eq:RR}
  \begin{bmatrix}
    n_{11}  & \dots & n_{1C} \\
    \vdots  &  \ddots & \vdots \\
    n_{S1}  & \dots & n_{SC} 
  \end{bmatrix}
  \begin{bmatrix}
    \varepsilon_1 \\
    \vdots \\
    \varepsilon_C
  \end{bmatrix}
  =
  \begin{bmatrix}
    E_1 \\
    \vdots \\
    E_S 
  \end{bmatrix}
  ,
\end{equation}
where $n_{Ic}$ is number of atoms in cluster $c$, for structure $I$, with a total of $S$ structures observed.
Eq.~(\ref{eq:RR}) can be restated as 
\begin{equation}
  \label{eq:OLS}
  \mathbf{X\varepsilon} = \mathbf{E},
\end{equation}
which ordinary least squares estimate the solution to by minimizing
\begin{equation}
  \label{eq:OLSM}
  \mathcal{E} = ||\mathbf{X\varepsilon - E}||^2,
\end{equation}
i.e.~the sum of squared residuals. Eq.~(\ref{eq:OLSM}) is minimized by
\begin{equation}
  \label{eq:OLSS}
  \mathbf{\varepsilon} = (\mathbf{X}^{\text{T}}\mathbf{X})^{-1}\mathbf{X}^{\text{T}}\mathbf{E}.
\end{equation}
However, depending on the rank of $\mathbf{X}$, $\mathbf{X}^{\text{T}}\mathbf{X}$ can potentially be singular. To overcome this problem ridge regression is used by altering Eq.~(\ref{eq:OLSM}) to
\begin{equation}
  \label{eq:RRM}
  \mathcal{E} = ||\mathbf{X\varepsilon - E}||^2 + \lambda ||\mathbf{\varepsilon}||^2,
\end{equation}
where $\lambda$ is a positive parameter chosen by the user. Eq.~(\ref{eq:RRM}) is minimized by 
\begin{equation}
  \label{eq:RRS}
  \mathbf{\varepsilon} = (\mathbf{X}^{\text{T}}\mathbf{X} + \lambda \mathbf{I})^{-1}\mathbf{X}^{\text{T}}\mathbf{E},
\end{equation}
where $\mathbf{I}$ is the identity matrix. Eq.~(\ref{eq:RRS}) always exists and has the added benefit of preventing overfitting by regularization on the free parameters $\mathbf{\varepsilon}$.
\\ \\
\subsection{Optimization}
\begin{figure}
  \centering
    \includegraphics[width=\columnwidth]{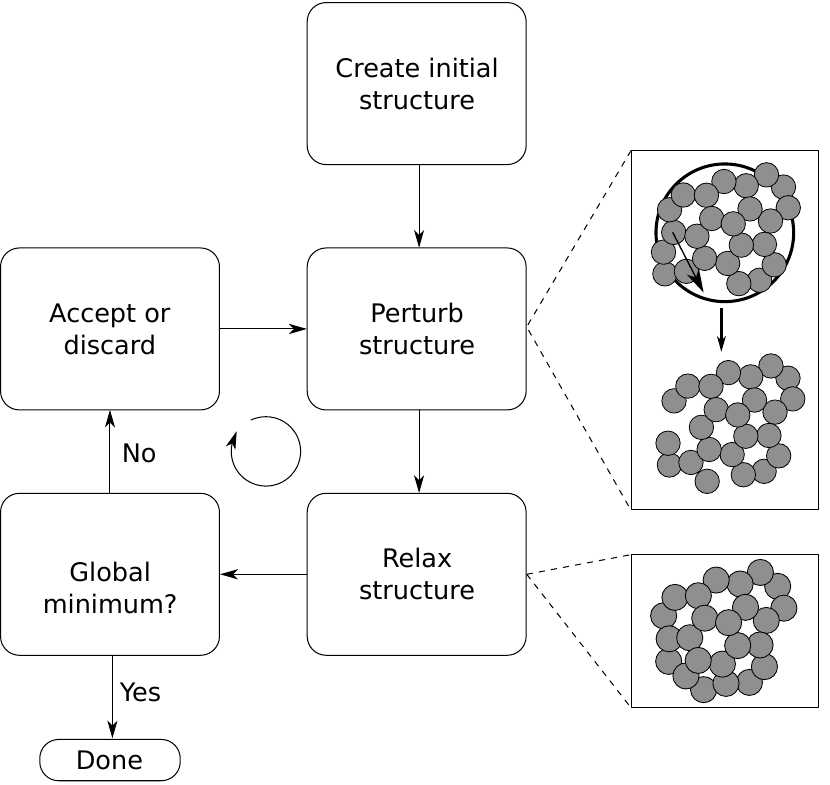}
  \caption{Illustration of the basin hopping framework.}
  \label{fig:BH}
\end{figure}

To demonstrate the applicability of the auto-bag feature and the local
energies we will derive and use these during global optimization runs. Fig.\ \ref{fig:BH} illustrates the layout of a BH search for the
global minimum energy structure of a set of atoms.

First, a random structure
is initiated and relaxed into the nearest local minimum energy structure according to the
atomic forces. Next, the structure undergoes some perturbative modification. Many strategies
may exist for this step. In the present context we have chosen a simple procedure that we shall
refer to as the \textit{fireworks perturbation}. With this procedure,
a random number, $N_{at}$, of atoms are repositioned uniformly within a disk centered at the center of mass of the structure (Fig.\ \ref{fig:BH}, black ring). The radius of the disk is a parameter analogous to the rattle distance in a rattle mutation. $N_{at}$ follows the normalized geometric series:
\begin{equation}
  \label{eq:gs}
\left.  P(x) = \left(\frac{1}{2}\right)^x \middle/ \sum_{x = 1}^{N} \left(\frac{1}{2}\right)^x \right.,
\end{equation}
where $N$ is the total number of atoms. With Eq.\ (\ref{eq:gs}), $P(1)\approx \frac{1}{2}$, $P(2)\approx \frac{1}{4}$, $\dots$, meaning that with about $50\%$ likelihood only one atom is
repositioned, with about $25\%$ likelihood two atoms are repositioned,
and so on. When the optimization is run without use of the local energies, the atoms are
chosen randomly. However, when the local energies are exploited, the $N_{at}$ atoms are drawn
one by one with a likelihood that also follows the normalized geometric series. That is, every time an atom is chosen, there will be about $50\%$ chance that it is the most unstable atom not yet chosen, about $25\%$ chance that it is the second most unstable and so on. A more elaborate expression could involve a dependence on the cluster energy and the number of clusters.

Once the structure has undergone the perturbation, it is relaxed according to the forces and a local minimum energy structure
is identified. This new structure replaces the previous structure according
to the Metropolis-Hastings criterion:
\begin{equation}
  \label{eq:mh}
  A =\min{\{1, \exp[\beta(E_{k - 1} - E_k)]\}}, 
\end{equation}
where $A$ is the probability of acceptance, $\beta = 1/k_B T$, $k_B$ is the Boltzmann constant and $T$ is a temperature parameter. Here $E_k$ is the potential energy of the newly found structure and $E_{k-1}$ is the potential energy of the previous structure. If the structure is accepted it serves as the starting point of the next perturbation, otherwise it is discarded.

\section{\label{sec:ljsystem}Lennard-Jones system}
As a first demonstration of the method, we consider a 2D structure of 19
atoms described by the classical Lennard-Jones (LJ) interaction
potential.  Using this simple potential has several benefits, it is
easily programmable meaning that the reader may code it and reproduce
our results. Local energies can be uniquely assigned to LJ atoms
meaning that the approximate machine learned local energies following
our method above may be benchmarked, and, importantly, calculating the
LJ potential is fast, which allows for fast testing and the production
of converged statistics on the efficiency.

\begin{figure}
  \centering
    \includegraphics[width=\columnwidth]{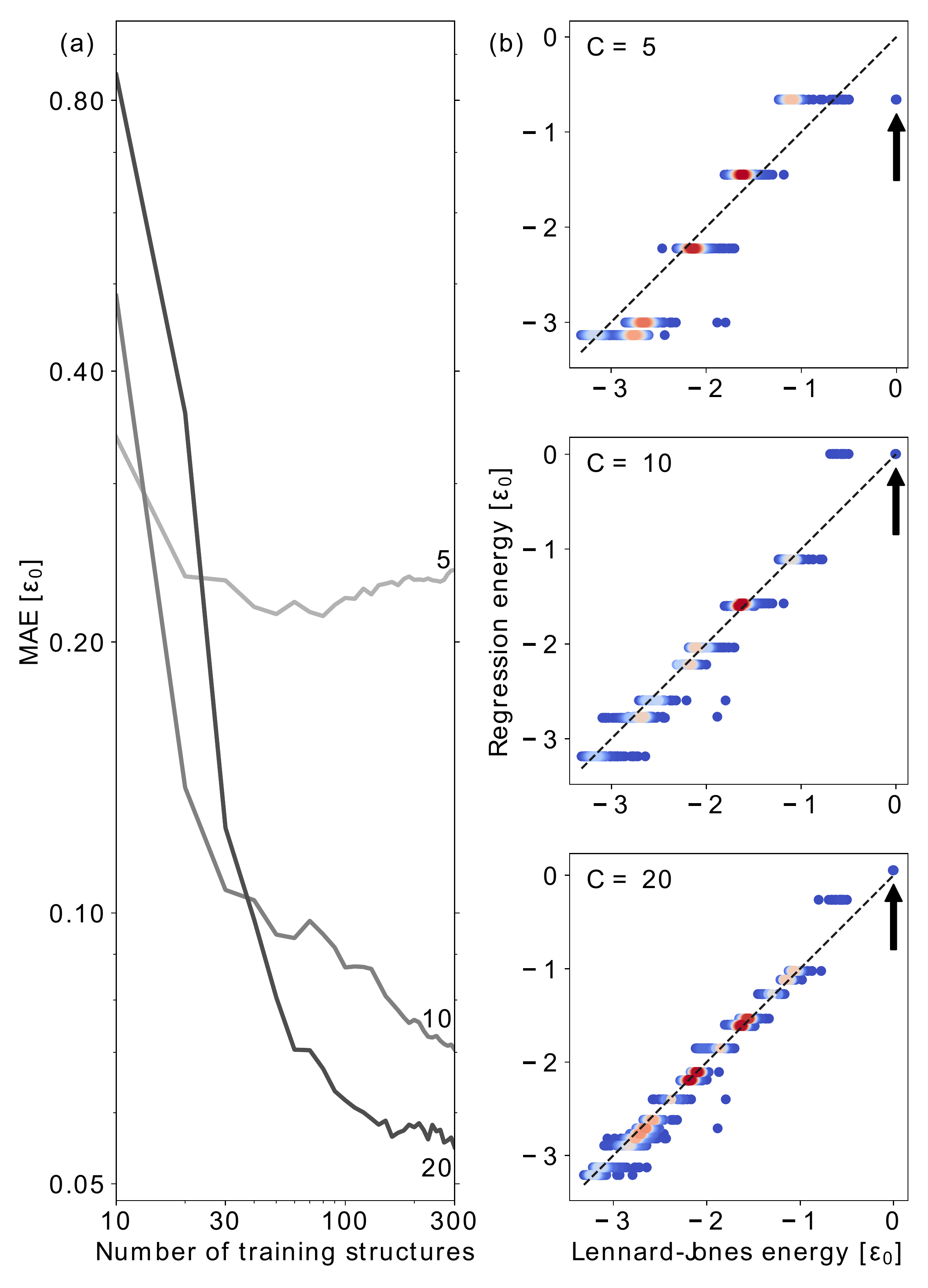}
  \caption{(a): Local energy learning curve for 19 atoms using varying number of clusters. The curve shows the mean absolute error (MAE) on the predicted energies of a set of test structures not available during training. (b): Prediction of local energies for unseen structures upon training on 300 configurations with 5, 10 or 20 clusters. The color indicates the density of data points. The three black arrows indicate atoms with no neighbors.}
  \label{fig:learningcurve}
\end{figure}

The LJ pair-potential is given by:
\begin{equation}
  \label{eq:lj}
  V(r) = \varepsilon_0 \left[\left(\frac{r_0}{r}\right)^{12} - 2\left(\frac{r_0}{r}\right)^6\right]. 
\end{equation}
$\varepsilon_0$ sets the energy scale and further turns out to be the depth of the well in the pair-potential, while $r_0$ sets the length-scale
and coincides with the equilibrium distance of the LJ dimer.
Upon training the model by solving Eq.~(\ref{eq:RRS}) for various data sizes and with varying number of clusters, the error on predicting local energies of structures not included in the training set is seen in Fig.\ \ref{fig:learningcurve}. The cutoff radius is chosen as $r_c = 2 \, r_0$ and a single radial Behler-Parrinello feature vector, Eq.~(\ref{eq:symfuncR}), with $r_s = 0$ and $\eta = 0.05$ is employed. Only relaxed structures are used for both training and testing, resulting in energies ranging from $-22\, \varepsilon_0$ to $-45 \, \varepsilon_0$. All relaxations and perturbations where confined to a plane causing the resulting structures to become strictly 2D.

Figure \ref{fig:learningcurve} shows that including more data in the training generally
leads to lower mean absolute errors (MAEs). However, with only few clusters, $C=5$, the
possible improvement of the MAE stagnates when utilizing around 100 different structures.
This behavior is as expected since restricting
the number of local environments will naturally provide a lower bound
on the error if less clusters than unique local environments are
used. Applying more clusters improves the lower bound on the error at
the expense of an increased error for small training samples due to
additional free parameters. Increasing the number of clusters prevents
dissimilar atoms from being forced into the same cluster and thus
leads to a more accurate energy prediction. This is seen in the
transition from 5 to 10 to 20 clusters where atoms not participating
in any chemical bonds (see arrows in Fig.\ \ref{fig:learningcurve})
initially belong to a non-zero-energy cluster, then a zero-energy cluster with non-zero-energy atoms and finally a zero-energy cluster
with only zero-energy atoms.

\begin{figure}
  \centering
    \includegraphics[width=\columnwidth]{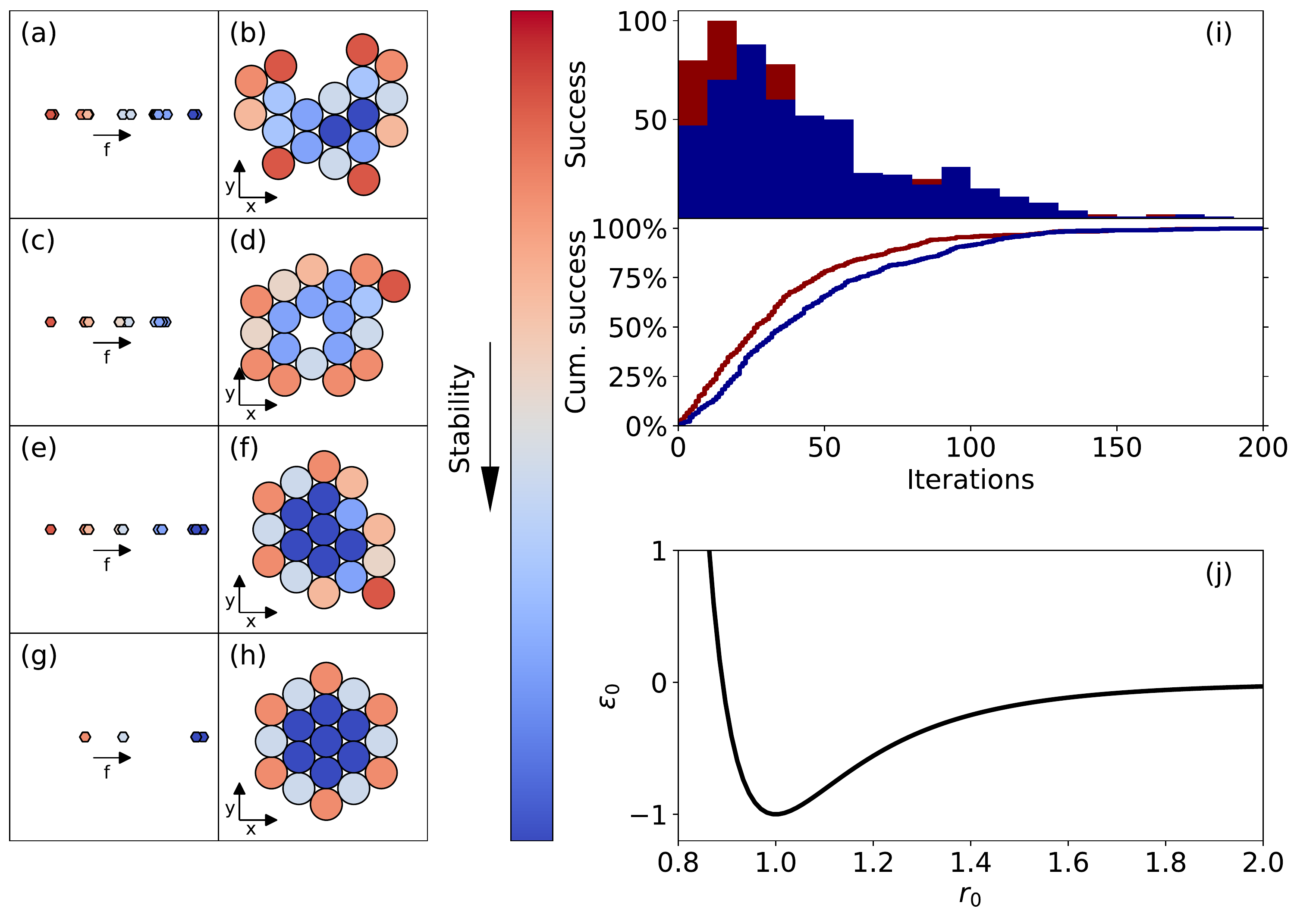}
  \caption{(a)-(h): Lennard-Jones structures and feature vectors colored to express the local energies based on ML model trained on 300 random structures. (i): Success curve for finding the global minimum without machine learning (blue) and with (red). (j): Lennard-Jones potential.}
  \label{fig:ljstrucs}
\end{figure}
A further inspection of the local
energies is seen in Fig.\ \ref{fig:ljstrucs} where several structures
have been colored according to predicted local energies from the fully
trained model with 10 clusters.
As the LJ energy correlates with the coordination of atoms and all atoms are placed approximately equidistantly to their nearest neighbors it is easily verified that the order of the local energies is correct for the global minimum shown show in Fig.\ \ref{fig:ljstrucs} (h). In Fig.\ \ref{fig:ljstrucs} (f) an indication of the applicability of local energies is seen as the only misplaced atom is predicted to be the most unstable. An inspection of the atom just above the most unstable atom reveals additional insight into the model as this atom, despite having four neighbors, is more unstable than the other atoms with four neighbors. The same tendency is observed in Fig.\ \ref{fig:ljstrucs} (b) and Fig.\ \ref{fig:ljstrucs} (d) where in all cases the most stable four-neighbor atom has three or more second-nearest neighbors, whereas the more unstable atom only has two. For atoms with two and five neighbors the same effect is observed. Thus, the ML model is able to correctly order very similar local environments.

Having shown that local energies are possible to learn it remains to be seen if they can be utilized in a optimization setting. Hence a ML model with 10 clusters is trained on-the-fly during a BH search and used to predict the local energies. 10 clusters are chosen based on optimizing the performance of the search as early as possible, while also obtaining an acceptable final error.
In each BH step, a number of atoms according to Eq.\ (\ref{eq:gs}) are perturbed. The atoms are chosen randomly in the benchmark run -- or dependent on their local energies according to the ML model as detailed in the discussion of Eq.\ (\ref{eq:gs}). The radius of the disk in the perturbation is $3 \, r_0$. Due to the small number of local minima for the LJ system the search is executed at $T = 0\text{K}$. 
From repeating the search 500 times, the cumulative success is seen in Fig.\ \ref{fig:ljstrucs} (i). We stress that each run starts with an untrained ML model. 
Only a minor increase in the success rate is observed, presumably due to the low complexity of the system. To test this hypothesis the search is repeated for a double well LJ system: \cite{DLJ}
\begin{equation}
  V(r) = \varepsilon_0 \left[\left(\frac{r_0}{r}\right)^{12} - 2\left(\frac{r_0}{r}\right)^6 - \exp{\Big(- \frac{(r-1.7r_0)^2}{2\sigma^2}\Big)}\right], 
\end{equation}
with $\sigma^2 = 0.02 \, r_0^2$. The global minimum is identical to that of the ordinary LJ potential and the same perturbation as before is used.
\begin{figure}
  \centering
   \includegraphics[width=\columnwidth]{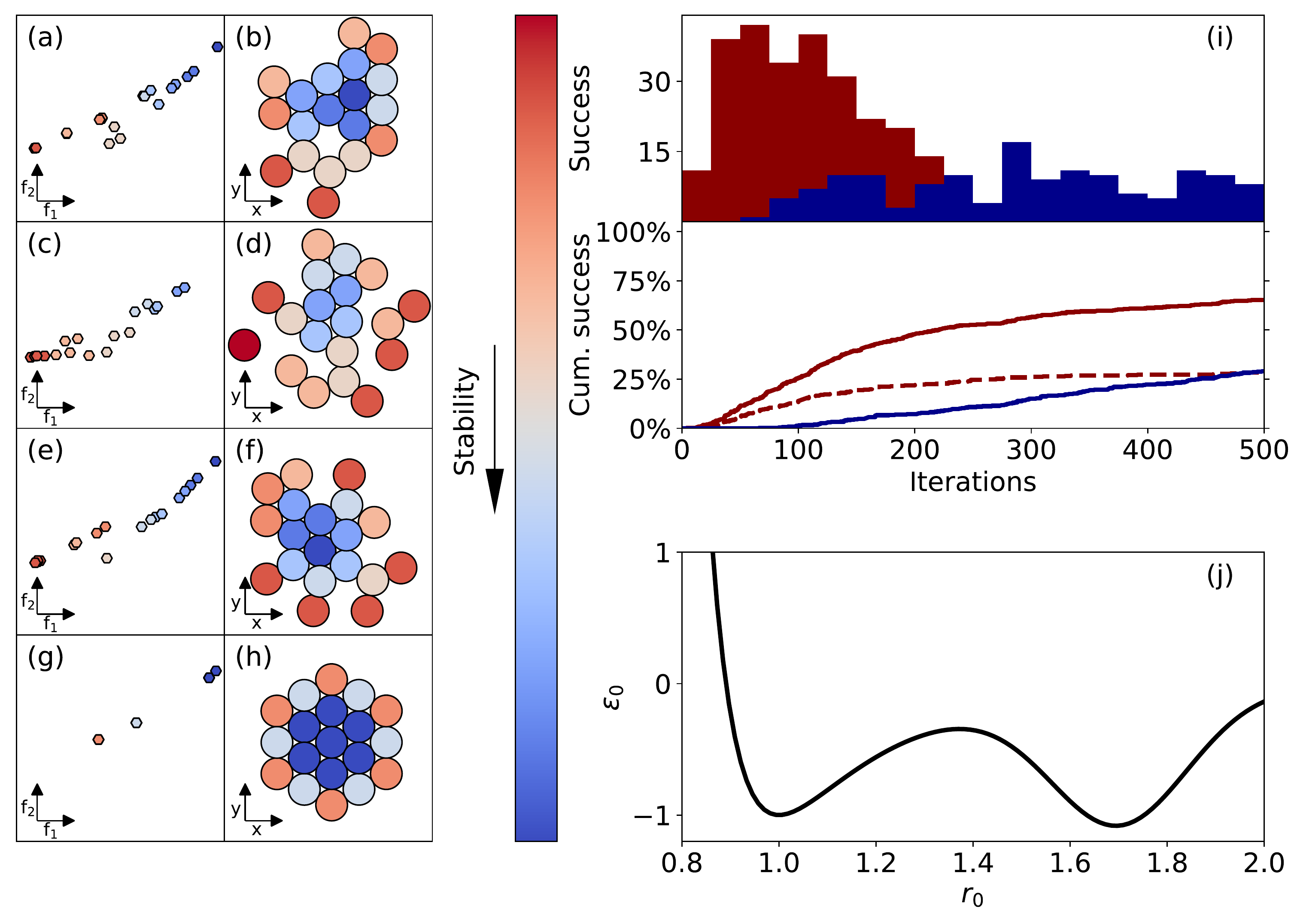}
   \caption{(a)-(h): Double-well Lennard-Jones structures and feature vectors colored to express the local energies based on a machine learning model trained on 300 random structures. (i): Success curve for finding the global minimum without machine learning (blue) and with 1D feature (Eq.~(\ref{eq:1dsym}), red dashed) and 2D feature (Eq.~(\ref{eq:2dsym}), solid red). (j): Double-well Lennard-Jones potential.}
  \label{fig:cumsuccess2}
\end{figure}
As the pair-potential has two minima, more structures may evolve as meta-stable local minima in configuration space, and as a consequence, finding the global minimum becomes a harder ordeal. As a measure to increase the success rate of finding the global minimum energy structure, the BH runs were performed at a finite temperature of $k_B T = 0.1 \, \varepsilon_0$. Yet, the search remains a challenge as testified to by Fig.\ \ref{fig:cumsuccess2} where the benchmark run now takes 500 BH iterations to find the global minimum with {\raise.17ex\hbox{$\scriptstyle\sim$}$20\%$ likelihood, while it took about 200 BH iterations to find it with almost $100\%$ certainty with the standard LJ pair-potential. 

As the 19-atom structure represents a harder problem with the
double well LJ pair-potential, it becomes easier to demonstrate the
beneficial effects of the ML approach introduced in this
work. The dashed red line of Fig.\ \ref{fig:cumsuccess2} shows how
greater success is achieved when local energies are derived based on a
one-dimensional feature vector and exploited in the BH
search. However, even more striking is the success rates achieved when a two-dimensional feature vector is employed, as shown by the solid red line in Fig.\ \ref{fig:cumsuccess2}. Now, the {\raise.17ex\hbox{$\scriptstyle\sim$}$20\%$ success level in finding the global minimum is achieved after a mere 100 BH iterations, representing a five-fold rate increase over the benchmark run. The two-dimensional feature vector contains
an angular component of the Behler-Parrinello type with parameters: $\eta = 0.005$, $\lambda = 1$ and $\xi = 1$.
Extending the feature vector with an angular component clearly
outperforms the one-dimensional one. We attribute this to a richer variety of local environments being present in the relaxed structures of the double well potential compared to the standard LJ potential. The increased performance seen with LJ type potentials motivates a search using a quantum mechanical energy expression as with DFT, where
the both the energy landscape and the local environments can be much
more complex.

\section{\label{sec:carbon}Density Functional Theory system}
Owing to the non-local Hamiltonian of quantum systems it is uncertain whether useful local energy information can be extracted. To investigate this we use DFT to describe a 3D system of 24 carbon-atoms utilizing a basis of linear combination of atomic orbitals for computational efficiency as available in the GPAW\cite{GPAW1, GPAW2} code with the Atomic Simulation Environment (ASE)\cite{ASE} package managing the atomic structures and optimization. To describe the exchange and correlation effects the PBE functional\cite{PBE} was chosen. The computational cell was constructed with no atoms closer than $6\, \text{\AA}$ to the non-periodic cell boundaries.
The optimization task is conducted using the auto-bag feature based on 10 clusters. The local feature vector is expanded to 13 dimensions using standard Behler-Parrinello symmetry functions with cutoff radius $r_c = 2 \, \text{\AA}$, $r_s = 0$, and default remaining parameters taken from Ref.\ [\onlinecite{AMP}] (see Table \ref{tab:param} in the appendix). The cutoff radius is chosen such that only nearest neighbors are included. This choice supports the extraction of the local energies early on in the global minimum searches where the datasets are small. Tests with larger cutoff showed a need for more data to learn useful local energies. 
To further prevent stagnation a parallel tempering scheme \cite{PT1} is employed where four BH searches at different temperatures are performed simultaneously. Every five iterations temperature swaps between simulations with adjacent temperatures $(i, j)$ are attempted and accepted with probability
\begin{equation}
  \label{eq:PT}
  A =\min{\{1, \exp[(\beta^i_k - \beta^j_k)(E^i_k - E^j_k)]\}}, 
\end{equation}
where $\beta^i_k = 1/k_B T^i_k$ and $E^i_k$ is the potential energy as before. The subscript refers to the structure at iteration $k$ and the superscript is an index on the parallel runs. Stagnated structures will then eventually acquire a higher temperature allowing them to escape local minima. Temperatures are chosen as $k_B T = [0.200 \, \text{eV}, 0.293 \, \text{eV}, 0.425 \, \text{eV}, 0.620 \, \text{eV}]$, keeping a constant ratio between adjacent temperatures as suggested in the literature\cite{PT3}. The temperatures are chosen to span both low temperatures allowing for exploitation as well as high temperatures for efficient exploration. The highest temperature is chosen to give a $20 \%$ chance of accepting a structure $1 \, \text{eV}$ higher than the current energy, and the lowest temperature such that almost only lower energy structures are accepted.
In order to evaluate the efficiency of the local energies a full parallel tempering minimum search is conducted. In this work we presume that the $\text{D}_{6h}$ structure is the global minimum. To direct our search towards planar structures the same perturbation as for the two-dimensional LJ system is used but with a disk of radius $4 \, \text{\AA}$. While the perturbation action was 2D, structural relaxation was done without any constraints leading to the structure becoming quasi-2D and occasionally 3D. As a benchmark the same parallel tempering algorithm is run with atoms to be perturbed chosen stochastically.

\begin{figure*}[tb]
  \centering
    \includegraphics[width=\textwidth]{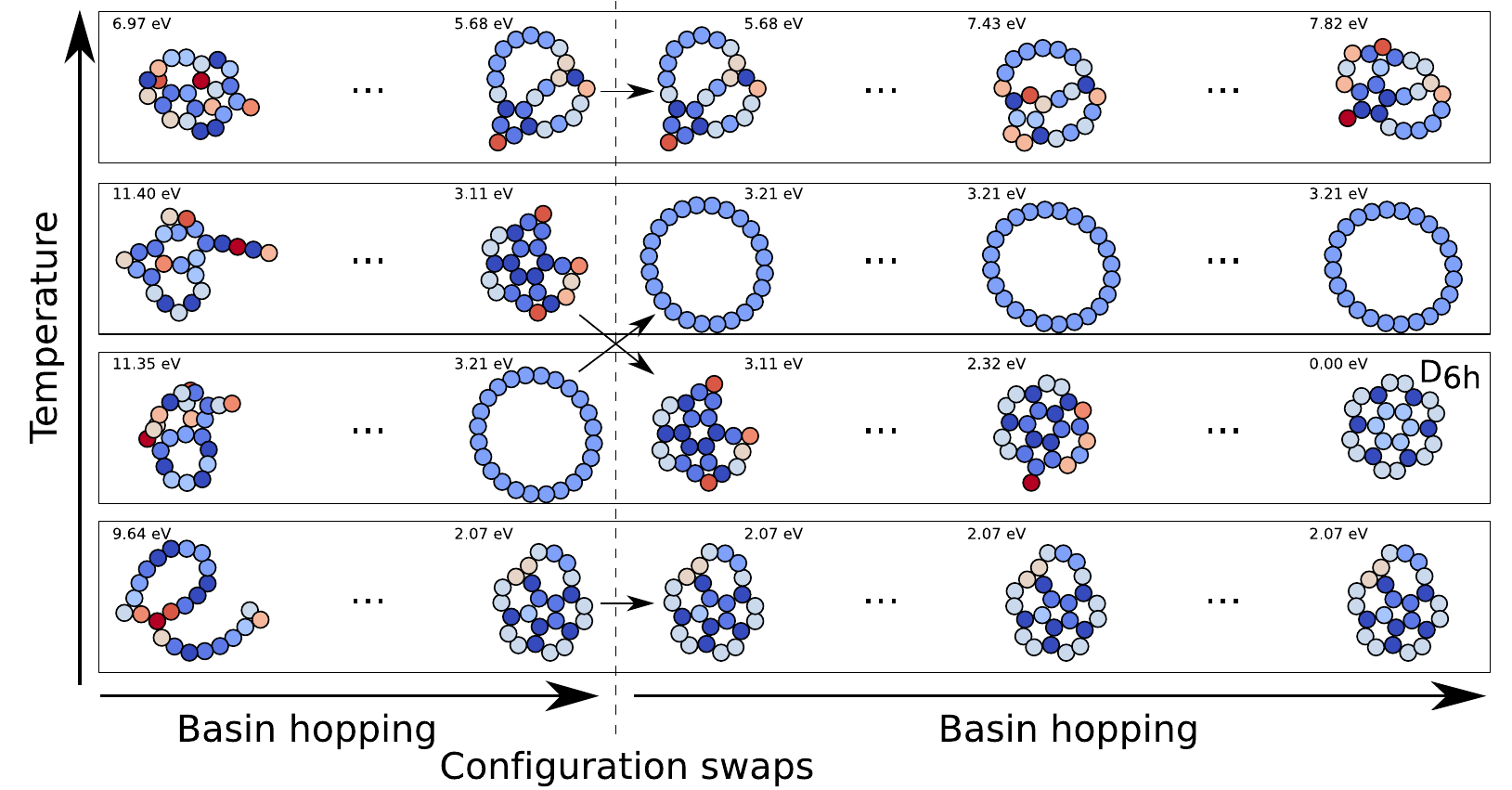}  
  \caption{Initial, intermediate and final structures found by a parallel tempering search with local energies depicted as the model predicted them on-the-fly. Configurational swaps were made possible for every five basin hopping iterations, but for clarity only one such swap is shown. Likewise, most basin hopping steps are omitted. The total length of the run was 90 iterations. DFT based total energies are given relative to the presumed global minimum.}
  \label{fig:run1}
\end{figure*}
In Fig.\ \ref{fig:run1} a parallel tempering run for C$_{24}$ structures with DFT potential energy is illustrated. Structures are shown at selected iterations, and the swapping action is shown at one given iteration. The structures and energies reported show how the highest temperature run remains agile and keeps exploring new structures through the run, while the lower temperature runs exploit found structures and perform refinements (or remain stuck). Eventually, the run with the second lowest temperature identifies the presumed global minimum energy structure ($\text{D}_{6h}$ with $E\equiv 0.00$ eV) and the calculation was stopped. Atoms are colored according to the ML model energy prediction at the given time of the search. Since the ML model
is refined on-the-fly, as more training data is accumulated, it is not
possible to compare colors for different iterations, especially not
for the initial and earliest iterations. However, in general it is
observed that atoms pointing out of the structures are drawn in red or
reddish colors meaning that they are the more unstable atoms. This is
for instance seen in the structure prior to the global minimum. Here
the unsaturated carbon atom is shown to be extremely unstable. In the
same structure it is also observed that one 5-membered ring and the 7-membered
ring contain unstable atoms, whereas 6-membered rings are shown to be
stable. The other 5-membered ring of that structure is composed of more stable atoms according to the modeled local energies.
This is an effect of an atom in this ring binding to the low coordinated high energy atom (colored dark red) and shows how
the auto-bag feature captures the details in the local environments of atoms.
\begin{figure}[]
  \centering
    \includegraphics[width=\columnwidth]{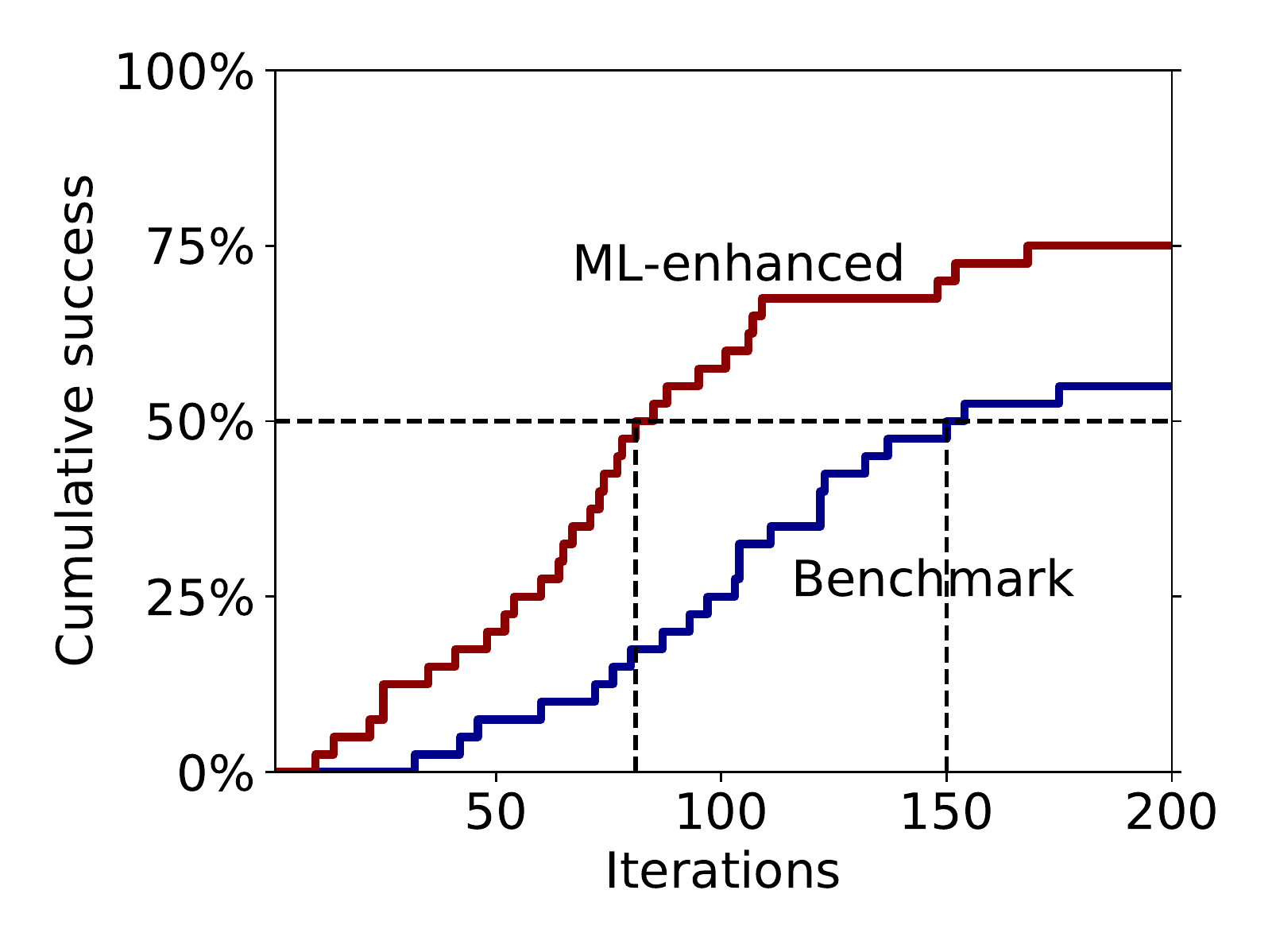}
  \caption{Cumulative success rate over 40 independent parallel tempering search runs for $\text{C}_{24}$ described with DFT. The blue curve represents the benchmark situation, where perturbed atoms are picked randomly. The red curve shows the results when perturbed atoms are picked according to their machine learned local energies, Eq.\ (\ref{eq:RRS}), using the auto-bag feature. Stagnation in fullerene type 3D structures more stable than the $\text{D}_{6h}$ 2D structure occurred rarely (2.5\% and 5\% for benchmark and machine learning runs, respectively) and was counted as failure.}
  \label{fig:cs}
\end{figure}

The cumulative success for 40 runs is seen in Fig.\ \ref{fig:cs}
displaying a convincing boost in the success rate for the ML enhanced
run. Note the improvement in cumulative success already in the
beginning of the run, demonstrating the limited amount of data
necessary to achieve reasonable local energies. In order to reach 50\%
success the ML enhanced approach required 81 attempts, which is 69
iterations less than the ordinary algorithm that required 150
iterations. Using the computationally inexpensive ML
model described in this work thus produced a speed-up factor of almost
two when applied to a full atomic-scale structural search within a DFT setting.

\section{\label{sec:conclusion}Summary}
In this work we have introduced the \textit{auto-bag} feature vector
that combines a local feature vector for each atom in a structure,
unsupervised machine learning (K-means) to establish clusters of such
local feature vectors, and supervised machine learning (ridge
regression) to extract atomic energies. The method was first
demonstrated to be capable of extracting the local energies for a
pair-wise classical potential of the Lennard-Jones form, where the
local energies are well-defined. Next, these machine learned local
energies were used to speed up the search for the global minimum
energy structure of 19 atoms described by the standard Lennard-Jones
potential or a more challenging double well Lennard-Jones
potential. Finally, the methodology was applied to a density functional theory, description of structures of 24 carbon atoms.  Here the local energies might be ill-defined, yet
our results show that the stochastic search for the global minimum
energy structure using the method of parallel tempering may be sped up
considerably when perturbing in the basin hopping steps preferentially
atoms that are predicted to be more unstable. The elements of the
method are rather simple and both Behler-Parrinello feature vectors as well as clustering have been shown to work for multi-component systems \cite{AMP, knud}. We thus expect similar behavior for structural searches within chemical physics in general including multi-component systems, molecules, nanoparticles, and solids. 

\begin{acknowledgments}
Grants from VILLUM FONDEN (Investigator grant, project number 16562) and the Danish Research Council (0602-02566B) have supported this research.
\end{acknowledgments}

\appendix
\section{Local feature vector}
To describe an atomic environment, we use the symmetry functions proposed by Behler and Parrinello\cite{BP}, which ensure rotational and translational invariance. The feature vector of atom $i$ is composed of pairwise- and triple-atom interactions given as
\begin{equation}
  \label{eq:symfuncR}
  f_i^\alpha(\mathbf{r}_1, \dots , \mathbf{r}_N) = \sum_{j \neq i} e^{-\eta(r_{ij} - r_s)^2/r_c^2}f_c(r_{ij}), 
\end{equation}
and
\begin{align}
  \label{eq:symfuncA}
  f_i^\beta(&\mathbf{r}_1, \dots , \mathbf{r}_N) = 2^{1 - \xi}  \sum_{j, k \neq i (j \neq k)} (1 - \lambda \cos{\phi_{ijk}})^{\xi} \nonumber \\ & \times e^{-\eta(r_{ij}^2 + r_{ik}^2 + r_{jk}^2)/r_c^2} f_c(r_{ij}) f_c(r_{ik}) f_c(r_{jk}) , 
\end{align}
respectively. Here $j$ and $k$ denote the index of other atoms, $r_{ij}$ is the distance between atom $i$ and atom $j$, $\phi_{ijk}$ is the valence angle between atom $i, \, j$ and $k$, centered at atom $i$ and $\eta, \, r_s, \, \xi$ and $ \lambda$ are parameters. By using multiple sets of parameters one can achieve a detailed description of the local environment, resulting in a high-dimensional local feature vector. See Ref.\ [\onlinecite{SYMFUNC}] for more information on choosing the parameters. The interactions are only accounted for within a sphere of radius $r_c$ by the use of the cutoff function
\begin{equation}
  \label{eq:cutoff}
  f_c(r) = \begin{cases} 0.5(1 + \cos{\pi r / r_c}), & \mbox{if } r \leq r_c \\ 0, & \mbox{if } r \geq r_c \end{cases}
\end{equation}
which is a smoothly decaying function approaching zero at $r = r_c$.
For the 13-dimensional feature vector the parameters can be seen in Table \ref{tab:param}.

\setlength{\tabcolsep}{10pt}
\begin{table}[h]
\centering
\caption{Behler-Parrinello parameters}
\label{tab:param}
\label{my-label}
\begin{tabular}{llllllllll}
  \hline
  \multicolumn{9}{c}{Radial symmetry functions} \\
  \hline
  $\eta$ & & 0.05 & 2 & 4 & 8 & 20 & 40 & 80 \\
  $r_s$ & & 0 & & & & & & \\
  \hline
  \multicolumn{9}{c}{} \\
  \multicolumn{9}{c}{} \\  
  \hline
  \multicolumn{9}{c}{Angular symmetry functions} \\
  \hline
  $\xi$ & & 1 & 2 & 4 & & & &  \\
  $\lambda$ & & 1 & -1 & & & & & \\
  $\eta$ & & 0.005 & & & & & & \\
  \hline
\end{tabular}
\end{table}

\bibliography{refs}

\end{document}